\documentclass[conference]{IEEEtran}
\IEEEoverridecommandlockouts
\usepackage{amsmath,amssymb,amsfonts}
\usepackage{textcomp}
\usepackage{xcolor}
\usepackage{amsmath,amsfonts}
\usepackage{algorithmic}
\usepackage{algorithm}
\usepackage{array}
\usepackage[caption=false,font=normalsize,labelfont=sf,textfont=sf]{subfig}
\usepackage{textcomp}
\usepackage{stfloats}
\usepackage{url}
\usepackage{verbatim}
\usepackage{graphicx}
\usepackage{caption}
\usepackage{romannum}
\usepackage{cite}
\usepackage{amsmath}
\usepackage{makecell}
\usepackage{multirow}
\usepackage{lipsum}
\usepackage{svg}
\def\BibTeX{{\rm B\kern-.05em{\sc i\kern-.025em b}\kern-.08em
    T\kern-.1667em\lower.7ex\hbox{E}\kern-.125emX}}
\begin{document}

\title{Deep Reinforcement Learning-Enabled Adaptive Forecasting-Aided State Estimation in Distribution Systems with Multi-Source Multi-Rate Data\\
\vspace{-10pt}
}


\author{Ying Zhang,
        Junbo Zhao,
        Di Shi,
        Sungjoo Chung 
\thanks{Y. Zhang and S. Chung are with the Department of Electrical and Computer Engineering, Montana State University, Bozeman, MT 59717, USA (e-mail: ying.zhang2@montana.edu; sungzhoo2@gmail.com).

J. Zhao is with the Department of Electrical and Computer Engineering, University of Connecticut, Storrs, CT 06269, USA (e-mail: junbo@uconn.edu). D. Shi is with the Department of Electrical and Computer Engineering, New Mexico State University, Las Cruces, NM 88003, USA (e-mail: dshi@nmsu.edu).
}}
\maketitle
\vspace{-10pt}
\begin{abstract}
Distribution system state estimation (DSSE) is paramount for effective state monitoring and control. However, stochastic outputs of renewables and asynchronous streaming of multi-rate measurements in practical systems largely degrade the estimation performance. This paper proposes a deep reinforcement learning (DRL)-enabled adaptive DSSE algorithm in unbalanced distribution systems, which tackles hybrid measurements with different time scales efficiently. We construct a three-step forecasting-aided state estimation framework, including DRL-based parameter identification, prediction, and state estimation, with multi-rate measurements incorporating limited synchrophasor data. Furthermore, a DRL-based adaptive parameter identification mechanism is embedded in the prediction step. As a novel attempt at utilizing DRL to enable DSSE adaptive to varying operating conditions, this method improves the prediction performance and further facilitates accurate state estimation. Case studies in two unbalanced feeders indicate that our method captures state variation with multi-source multi-rate data efficiently, outperforming the traditional methods.
\end{abstract}


\section{Introduction}

Distribution system state estimation (DSSE) relates redundant meter readings to network states for advanced system situational awareness \cite{1primadianto2017review}. DSSE demands more affordable solutions since distribution systems have fundamental differences from transmission systems, such as low measurement redundancy, high $\mathrm{r} / \mathrm{x}$ ratios, and unbalanced operation \cite{3pau2013efficient}.

This paper focuses on DSSE with multi-source multi-rate hybrid measurement data in unbalanced distribution systems with penetration of distributed energy resources (DERs). Notably, these measurements installed in the systems differ in time resolution and amount, and the existence of slow-rate data renders the distribution system often unobservable within a short timescale. For instance, measurements collected by supervisory control and data acquisition (SCADA) systems and short-term forecasting of load consumption and DERs as pseudo-measurements are usually obtained at minute-level time resolution \cite{14alcaide2018electric}. Compared with the meter readings, these pseudo-measurements tend to be of poor quality.  In contrast, phasor measurement units (PMUs) record synchrophasors of voltages tens of readings per second \cite{5vonmeier2017precision}. However, the number of PMUs is limited and cannot meet the distribution system observability independently \cite{4zhang2020towards}. Due to the volatility of load and high penetration of DERs, the capability of DSSE running for real-time monitoring and control is constrained immensely by these slow-rate measurements. 

Most approaches combining multi-source measurements in distribution systems fall into the category of Kalman filters (KFs)
\cite{11carquex2018state,10huang2015evaluation}. For example, \cite{11carquex2018state} and \cite{10huang2015evaluation} utilize previous estimates to improve estimation accuracy by ensemble KFs and extended KFs, respectively. However, their performance in unbalanced distribution systems is not investigated. 
Several recent efforts leverage fast time resolution and high precision of PMUs to improve the accuracy of DSSE. \cite{12sarri2012state} uses a high measurement redundancy and assumes that the number of SCADA meters is sufficient and solely satisfies observability requirements. Nevertheless, this assumption is not practical in distribution feeders. \cite{13song2020dynamic} formulates a first-order prediction-correction optimization model using multi-rate PMU and pseudo-measurement data. Besides,  \cite{10huang2015evaluation,11carquex2018state,12sarri2012state,13song2020dynamic} do not consider the impact of DER penetration. The underlying high DER volatility might result in frequent time-varying system states \cite{VR_PV} and make these algorithms hard to obtain efficient estimation. Recently, historical data stored by the system operators is utilized to develop data-driven methodologies for the fusion of multi-source measurements in distribution systems \cite{Bayesian9723534}. To this end, Zhao et al. \cite{Zhao9087787} forecasted the data of load and DERs injections through the support vector machine and propose a robust state estimation algorithm. A nonlinear autoregressive Gaussian process approach is proposed in \cite{Zhao10184461} for voltage probabilistic estimation. However, the asynchronorization issue of multi-source measurements (i.e., PMUs and other types), as abovementioned, is not fully addressed in the literature.

To fill the gap, we propose a deep reinforcement learning (DRL)-enabled adaptive forecasting-aided state estimation (FASE) algorithm using multi-source multi-rate data in unbalanced distribution systems. A three-step FASE framework with multi-source multi-rate data is proposed, including DRL-aided parameter identification, prediction, and state estimation steps. In particular, we develop a data-driven DRL algorithm by offline training for adaptive parameter identification prior to the prediction step, where multi-rate historical data stored by system operators are leveraged to optimize the prediction step. DRL, to the best of our knowledge, is here applied to fuse multi-source multi-rate data and enhance the FASE accuracy for the first time. 
The main contributions include 1)
a novel adaptive FASE algorithm to handle hybrid multi-rate measurements from limited PMU and slow-rate measurements and 2) superior estimation performances over conventional methods, illustrated by case studies in unbalanced distribution systems. 


\section{DSSE With Multi-Rate MeasurementS}

\subsection{Extended Kalman Filter for State Estimation}
Conventionally, the mathematical model for state estimation consists of process equations and measurement equations, expressed as:
\begin{equation}
\mathbf{x}_{k+1}=f\left(\mathbf{x}_k\right)+\mathbf{\omega}_k , \\\ \mathbf{\omega}_k \sim \mathcal{N}(\mathbf{0}, \mathbf{Q})
\end{equation}
\begin{equation}
\mathbf{y}_k=h\left(\mathbf{x}_k\right)+\mathbf{v}_k,\\\ \mathbf{v}_k \sim \mathcal{N}(\mathbf{0}, \mathbf{R})
\end{equation}
where $\mathbf{x}_{k+1} \in \mathbb R^n$ and $\mathbf{x}_k\in \mathbb R^n$ denote the system state at time $k+1$ and $k$ respectively; $ f(\cdot)$ is the process function in this system, while $ h(\cdot)$ are the measurement function; $\omega_k$ is the noise vector of the process model; $\mathbf{y}_k\in \mathbb R^{m}$ denote the measurement vector, $\mathbf v_k$ denotes the measurement error.

We adopt a widely used extended KF algorithm in this paper to propose an adaptive FASE algorithm. To some extent, extended KF is capable of depicting the state change accurately when formulating the distribution system operation as a linearized model, to name a few in a wealth of literature, \cite{3pau2013efficient,4zhang2020towards}. Here we simply present the equations in the extended KF for the sake of brevity:

1) \textit{Prediction:} the state prediction and the covariance matrix of the error are computed using the known knowledge of $\mathbf{F}$:
\begin{equation}\label{statePre}
\hat{\mathbf{x}}_{t \mid t-1}=\mathbf{F} \hat{\mathbf{x}}_{{t-1}|{t-1}}+\mathbf G
\end{equation}
\begin{equation}
\boldsymbol{\Sigma}_{t \mid t-1}=\mathbf{F} \boldsymbol{\Sigma}_{t-1 \mid t-1}  \mathbf{F}^{\top}+\mathbf{Q}
\end{equation}
where $\boldsymbol{\Sigma}_{t-1 \mid t-1}$ denotes the covariance at the last time.

The observations $\mathbf y$ and innovation $\Delta \mathbf y$ are updated through the Jacobian matrix $\mathbf{H}$ as
\begin{equation}
\hat{\mathbf{y}}_{t|t-1}=\mathbf{H} \hat{\mathbf{x}}_{t|{t-1}} 
\end{equation}
\begin{equation}
\mathbf{S}_{t|t-1}=\mathbf{H}  \boldsymbol{\Sigma}_{t|t-1} \mathbf{H}^{\top}+\mathbf{R} 
\end{equation}
\begin{equation}
\Delta \mathbf{y}_t=\mathbf{y}_t-\hat{\mathbf{y}}_{t|t-1}=\mathbf{y}_t-\mathbf{H}\hat{\mathbf{x}}_{t|t-1}
\end{equation}

2) \textit{Update:} The \textit{a posteriori} state moments are computed based on the \textit{a priori} moments as
\begin{equation}\label{corr}
\hat{\mathbf{x}}_{t \mid t}=\hat{\mathbf{x}}_{t \mid t-1}+\mathcal{K}_t \Delta \mathbf{y}_t 
\end{equation}
\begin{equation}
\Sigma_{t \mid t}=\boldsymbol{\Sigma}_{t \mid t-1}-\mathcal{K}_t \mathbf{S}_{t \mid t-1}  \mathcal{K}_t^{\top} 
\end{equation}
\begin{equation}
\mathcal{K}_t=\boldsymbol{\Sigma}_{t|{t-1}} \mathbf{H}^{\top}  \mathbf{S}_{ t|{t-1}}^{-1} 
\end{equation}
where $\mathcal{K}_t $ is the gain matrix of KF, i.e., Kalman gain (KG). 
\begin{figure}[!t]
\centerline{\includegraphics[width=3.3in]{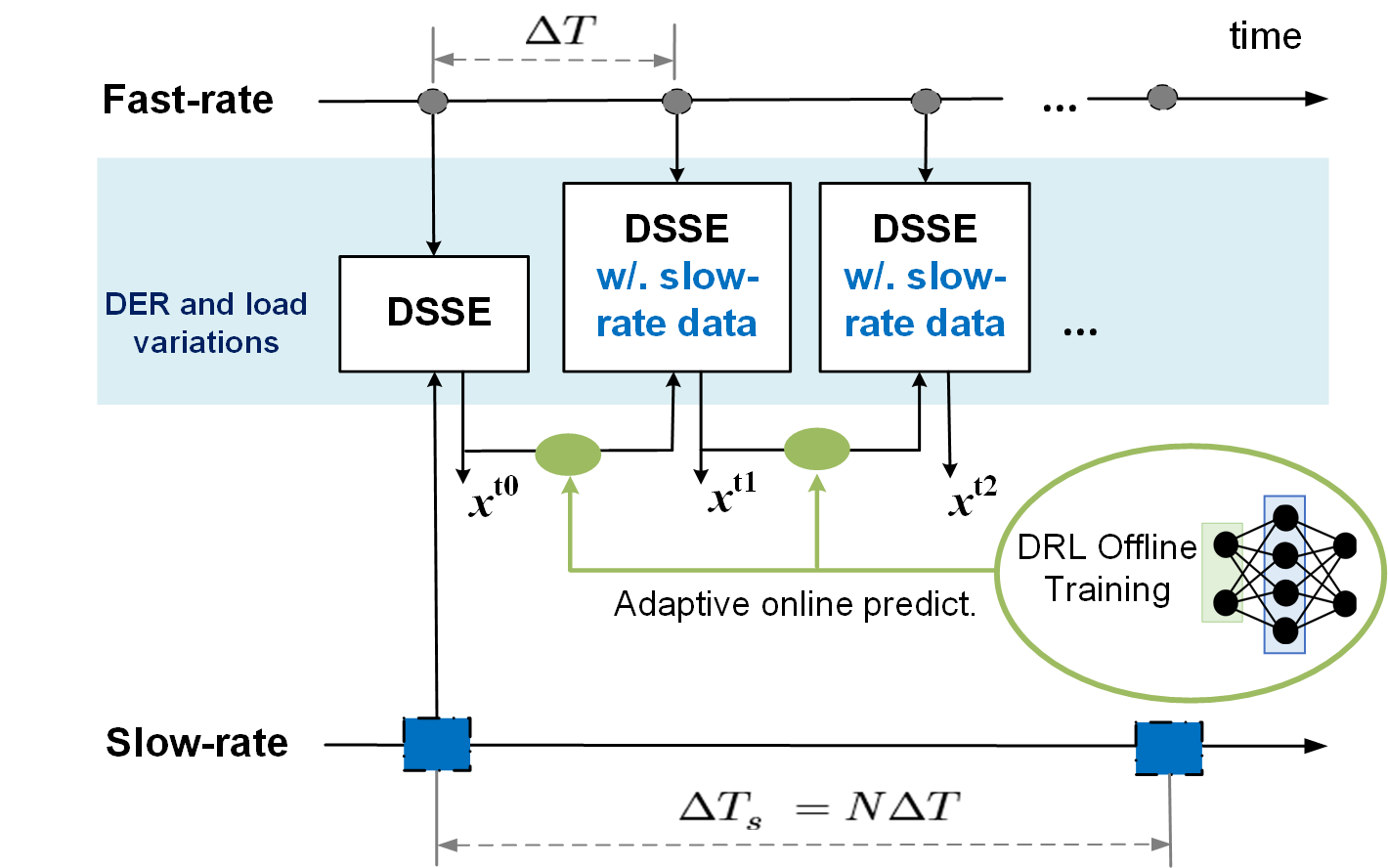}}
\caption{Proposed DRL-enabled adaptive DSSE framework.}
\label{fig1}
\vspace{-10pt}
\end{figure}
\subsection{Integration of Fast- and Slow-Rate Sensor Data}
In light of industry practice, we suppose the measurement dataset is composed of PMU phasors, pseudo-measurements, and SCADA data (smart meters can edge out pseudo-measurements as the feeder infrastructure gets upgraded). Conventional state estimators assume that all types of measurements are collected at the same snapshot. When running state estimation, the PMU data can be always available due to their high resolution. However, measurements of the slow-rate sensors, such as the SCADA system, are collected lagging far behind PMUs. We denote the run period of DSSE as $\Delta T$, and the reporting period of slow-rate measurements as $\Delta T_{s}$. If state estimation is expected to run for system monitoring within minutes to reflect the impact pf DER generation, we suppose $\Delta T_{s}=N\Delta T$ exists. Usually, $\Delta T_{s}$ and $\Delta T$ are specified by system operators according to operational practice, e.g., $\Delta T=6$s-$10$ min \cite{11carquex2018state}, while $\Delta T_s=60$ min \cite{10huang2015evaluation}. 

Since pseudo-measurements already feature very poor quality (with errors ranging from 10\%-50\%) \cite{1primadianto2017review}, we think the time skew issue between different types of sensors/sources except PMUs does not impact the relativity of high error covariance of these pseudo-measurements largely with other types. Hence, they can be reduced to slow-rate data with the same rate as the SCADA data if this kind of time skew issue indeed exists. 
In the proposed DSSE method, at a given time $t$, the measurement dataset consists of fast-rate PMU and other slow-rate measurements. Let 
$m_{s}$ and $m_{p}$ denote the number of data points of these two different rates in the measurement vector $\mathbf z_k$, and $m_p<n$ renders this system unobservable at the time step that only PMUs report.
\vspace{-5pt}
\section{Propose DRL-Enabled FASE Algorithm}

We propose an adaptive DRL-enabled FASE algorithm with multi-source multi-rate measurements, which is a three-step filter procedure, including DRL for parameter identification, state and slow-rate measurement prediction, and state estimation. Fig.1 depicts the framework of the proposed adaptive method. We predict the slow-rate measurements based on these previous estimations and restore the system observability at each $t$. Specifically, Holt's two-parameter exponential approach is used to build a recursive DRL-enabled prediction method, producing an adaptive FASE algorithm.

\subsection{Data-Driven Prediction with Slow-Rate Measurements}

Typical FASE includes a prediction step of system states and an update step. It is generally built on the assumption that prediction and measurement errors follow known Gaussian distributions. Nevertheless, due to the existence of slow-rate measurements, the statistical knowledge of $\mathbf v_t$ is no longer accurate. To address this, we propose an adaptive prediction model to maintain and leverage the statistical knowledge regarding these slow-rate measurements.



In Holt's two-parameter exponential approach, the prediction of the state at $t$ is attained by the state prediction and estimation in the previous time as \cite{14alcaide2018electric}:
\begin{equation} \label{Pred}
\begin{cases} 
    \hat{\mathbf{x}}_{t \mid t-1}=\mathbf{a}_{t-1}+\mathbf{b}_{t-1}\\
\mathbf{a}_{t-1}=\alpha_{t} \hat{\mathbf{x}}_{t-1}+\left(1-\alpha_{t}\right) \tilde{\mathbf{x}}_{t-1}\\
\mathbf{b}_{t-1}=\beta_{t}\left(\mathbf{a}_{t-1}-\mathbf{a}_{t-2}\right)+\left(1-\beta_{t}\right) \mathbf{b}_{t-2}
\end{cases}
\end{equation}
where $\alpha_{t}$ and $\beta_{t}$ denote smoothing coefficients, $\alpha_{t}, \beta_{t} \in[0,1]$.

Rewrite the prediction equation \eqref{Pred}, and we obtain:
\begin{equation}\label{Pred2}
\begin{cases} 
\mathbf F=\alpha_{t}\left(1+\beta_{t}\right)\mathbf I \\
\mathbf{G}=\left(1+\beta_{t}\right)\left(1-\alpha_{t}\right) \tilde{\mathbf{x}}_{t-1}-\beta_{t} \boldsymbol{a}_{t-2}+
\left(1-\beta_{t}\right) \mathbf{b}_{t-2}
\end{cases}
\end{equation}


The Holt's exponential prediction model can be gradually closer to the actual nonlinear DSSE model. Traditionally, the smoothing parameters can be determined by empirical knowledge or heuristics\cite{14alcaide2018electric}. However, factoring in the time-varying impacts of slow-rate measurements as the FASE method progresses, the smoothing coefficients are no longer constant and in biased accordance with the prior values. Thus, we denote them as $\mathbf{c}_{t}=\left[\mathbf{\alpha}_{t}, \mathbf{\beta}_{t}\right]^{\top}$, which is time-senstive and performs as extra auxiliary variables. 
To correct the smoothing coefficients $\mathbf{c}_{t}$ adaptively, we formulate an NN-based DRL agent by observing the state estimation and prediction at the last time step as well as the latest PMU data:
\begin{equation}\label{Gnn}
    \mathbf{c}_{t}=\mathcal G_{nn}(\hat{\mathbf x}_{t-1},\tilde{\mathbf x}_{t-1},\mathbf p_{t})
\end{equation}
where $\mathbf p_{t}$ denotes the available PMU measurements at time $t$.


Then, the slow-rate measurements, denoted as $\hat{\mathbf{d}}_{t \mid t-1}$, are reconstructed by the measurement function with respect to the latest predicted states, $\hat{\mathbf{x}}_{t \mid t-1}$:
\begin{equation}\label{slowpre}
    \hat{\mathbf{d}}_{t \mid t-1}=h_{s}\left(\hat{\mathbf{x}}_{t \mid t-1}\right)
\end{equation}
where $h_{s}(\cdot)$ denotes the measurement function of the slow-rate measurements.

Ideally, if sufficient historical data can be fully leveraged for the appropriate learning on the smoothing coefficients, the DRL learning agents have the potential in reconstructing predicted data regarding the slow-rate measurements with similar or close errors to the known statistics at the previous step, i.e., $\mathbf R$. In this way, the negative impact of the time skew issue from slow-rate measurements can be mitigated. Then, the validity of the previously known covariances for these slow-rate measurement errors is secured in the FASE procedure.

Therefore, integrating the measurement prediction in \eqref{slowpre} and real-time PMU data at time $t$, the relationships between the states $\mathbf{x}_{t}$ and measurements are expressed as:
\begin{equation}
\left[\begin{array}{c}
\mathbf{p}_{t} \\
\hat{\mathbf{d}}_{t \mid t-1}
\end{array}\right]=h\left(\mathbf{x}_{t}\right)+\mathbf{v}_{t} ,\\\ \mathbf{v}_t \sim \mathcal{N}(\mathbf{0}, \mathbf{R})
\end{equation}

Next, we elaborate on the design of the NN-based agent $\mathcal G_{nn}$ and propose a DRL-enabled adaptive FASE approach.
\subsection{DRL-Enabled Adaptive Forecasting-Aided State Estimation}
We propose a novel DRL-enabled adaptive parameter identification method in the prediction step of FASE. This method selects $\alpha_t$ and $\beta_t$ in \eqref{Pred}, and it is implemented online after offline training on historical multi-rate measurement data. It runs at intermediate steps (totally $N$) of two DSSE with complete measurement sets, $t=T_{\text {begin }}: \Delta T: T_{\text {end }}$, where $T_{\text {begin }}$ and $T_{\text {end }}$ are the beginning and end time of this identification procedure. Here $N$ is the number of time steps between two updates of slow-rate measurements

The parameter identification can be described as a $N$-step Markov decision process (MDP). In the decision making of $\alpha_t$ and $\beta_t$, MDP is defined by the tuple $(\mathcal{S}, \mathcal{A}, p, r)$, where $\mathcal{S}$ and $\mathcal{A}$ represent the state space and action space, and $p$ is an unknown state transition probability, $\mathcal{S} \times \mathcal{S} \times \mathcal{A}$, of the next state $\mathbf{s}_{t+1} \in \mathcal{S}$, given the current state $\mathbf{s}_{t} \in \mathcal{S}$ and action $\mathbf{a}_{t} \in \mathcal{A}$. One time step later, the new states become $\mathbf{s}_{t+1}$ according to $p_{\pi}\left(\mathbf{s}_{t}, \mathbf{a}_{t}\right)$. 


NNs as DRL agents interact with the environment, i.e., the FASE procedure, to decide the actions per $\Delta T$. The actions at $t$ are the smoothing parameters in the identification step of FASE and are denoted as $\mathbf c_t$, bounded within [0,1].
DRL here attempts to help FASE estimate the closest voltages to those of static DSSE at $t$ with perfect synchronization by tuning $\alpha_{t}$ and $\beta_{t}$ on each intermediate time step. The observation vector of the agent $\mathcal{G}_{nn}$ at $t$ is defined as:
\begin{equation}
    \mathbf{s}_{t}=\{\hat{\mathbf x}_{t-1},\tilde{\mathbf x}_{t-1}, \mathbf p_{t}\}
\end{equation}
where $\mathbf{s}_{t} \in \mathbb{R}^{2n+m_p}$. 

The core of DRL is to search for a decision policy $\pi$ that maximizes the sum of rewards at all the time steps in each training episode by
\begin{equation}
    \max _{\pi} \sum_{t=T_{\text {begin }}}^{T_{\text {end }}} \mathbb{E}_{\left(\mathbf{s}_{t}, \mathbf{a}_{t}\right) \sim p_{\pi}} r\left(\mathbf{s}_{t}, \mathbf{c}_{t}\right)
\end{equation}

We define the sum of the squared difference between the state prediction and estimation solutions as the reward function at $t$:
\begin{equation}
    r\left(\mathbf{s}_{t}, \mathbf{c}_{t}\right)=-\left(\tilde{\mathbf{x}}_{t}-\hat{\mathbf{x}}_{t}\right)^{\top}\left(\tilde{\mathbf{x}}_{t}-\hat{\mathbf{x}}_{t}\right)
\end{equation}

The closer to zero $r\left(\mathbf{s}_{t}, \mathbf{c}_{t}\right)$ is, the more similar estimation performance to DSSE with synchronized hybrid data is obtained by the proposed DRL-enabled adaptive FASE.

Here the proposed DRL framework is conducted by deep Q network (DQN) \cite{26zhang2021deep} for illustration, and the actions take values in $[0,1]$ at a step of 0.1. The NN-based agent updates an action-reward $Q$ function via the Bellman equation:
\begin{equation}\label{Bellman}
    Q\left(\mathbf{s}_{t+1}, \mathbf{c}_{t+1}\right)=Q\left(\mathbf{s}_{t}, \mathbf{c}_{t}\right)+\alpha\left(\hat{Q}\left(\mathbf{s}_{t}, \mathbf{c}_{t}\right)-Q\left(\mathbf{s}_{t}, \mathbf{c}_{t}\right)\right)
\end{equation}
where $\alpha$ denotes a learning rate and the $\operatorname{target} Q$ function $\hat{Q}\left(\mathbf{s}_{t}, \mathbf{c}_{t}\right)=r\left(\mathbf{s}_{t}, \mathbf{c}_{t}\right)+\gamma \cdot \max Q\left(\mathbf{s}_{t+1}, \mathbf{c}_{t+1}\right), \gamma \in[0,1]$.

Stochastic gradient descent on NN parameters $\boldsymbol{\theta}$ is adopted to minimize the following loss function, which enforces \eqref{Bellman}:
\begin{equation}
    \mathcal{L}(\boldsymbol{\theta})=\mathbb{E}\left[\frac{1}{2}\left(Q\left(\mathbf{s}_{t}, \mathbf{c}_{t}\right)-\hat{Q}\left(\mathbf{s}_{t}, \mathbf{c}_{t}\right)\right)^{2}\right]
\end{equation}

Experience replay and $\varepsilon$-greedy policy techniques are used in the training process. 
The detailed DRL training process and the proof of the convergence of the DRL for MDP can be found in \cite{26zhang2021deep} and its references, and we omit them here. In online tests, the smoothing parameters as the actions are provided by the well-trained agents for the prediction step.

The proposed method is summarized here. Based on the previous estimation results, predict the slow-rate measurements that are outdated at $t$; then, with real-time incoming PMU data, capture the latest system states and use them for the prediction at the next time step. The pseudo-code of the proposed algorithm is shown in Algorithm 1. 
\begin{algorithm}[H]
\caption{DRL-Enabled Adaptive FASE}\label{alg:alg1}
\begin{algorithmic}
\STATE {\textbf{Inputs:} Multi-source multi-rate data, distribution system model ${h}({\cdot})$, and time steps $t$, well-trained DRL agent $\mathcal G_{nn}$}. 
\STATE \textbf{for} each time step in $t=T_{\text {begin }}: \Delta T: T_{\text {end }}$ \textbf{do}
\STATE \hspace{0.5cm} \textbf{if} all the data is synchronized
\STATE \hspace{1.0cm} Solve the static DSSE model.
\STATE \hspace{0.5cm} \textbf{else}  
\STATE \hspace{1.0cm} \textbf{DRL-Based Parameter Identification:}
\STATE \hspace{1.0cm} Obtain smoothing coefficients $\alpha_{t}$ and $\beta_{t}$ in \eqref{Pred2}
\STATE \hspace{1.0cm} by the online implementation of the DRL agent.
\STATE \hspace{1.0cm} \textbf{Slow-rate Measurement Prediction:}
\STATE \hspace{1.0cm} Predict the states and slow-rate measurement data
\STATE \hspace{1.0cm} by \eqref{statePre} and \eqref{slowpre}.
\STATE \hspace{1.0cm} \textbf{State Estimation:}
\STATE \hspace{1.0cm} Update the states by \eqref{corr}. 
\STATE \hspace{0.5cm} \textbf{end if}
\STATE \textbf{Outputs:} $\hat{\mathbf{x}}_{t}$ 
\STATE \textbf{end for}
\end{algorithmic}
\label{alg1}
\end{algorithm}


\section{Case Study}

\begin{figure}[!t]
\centerline{\includegraphics[width=3.4in]{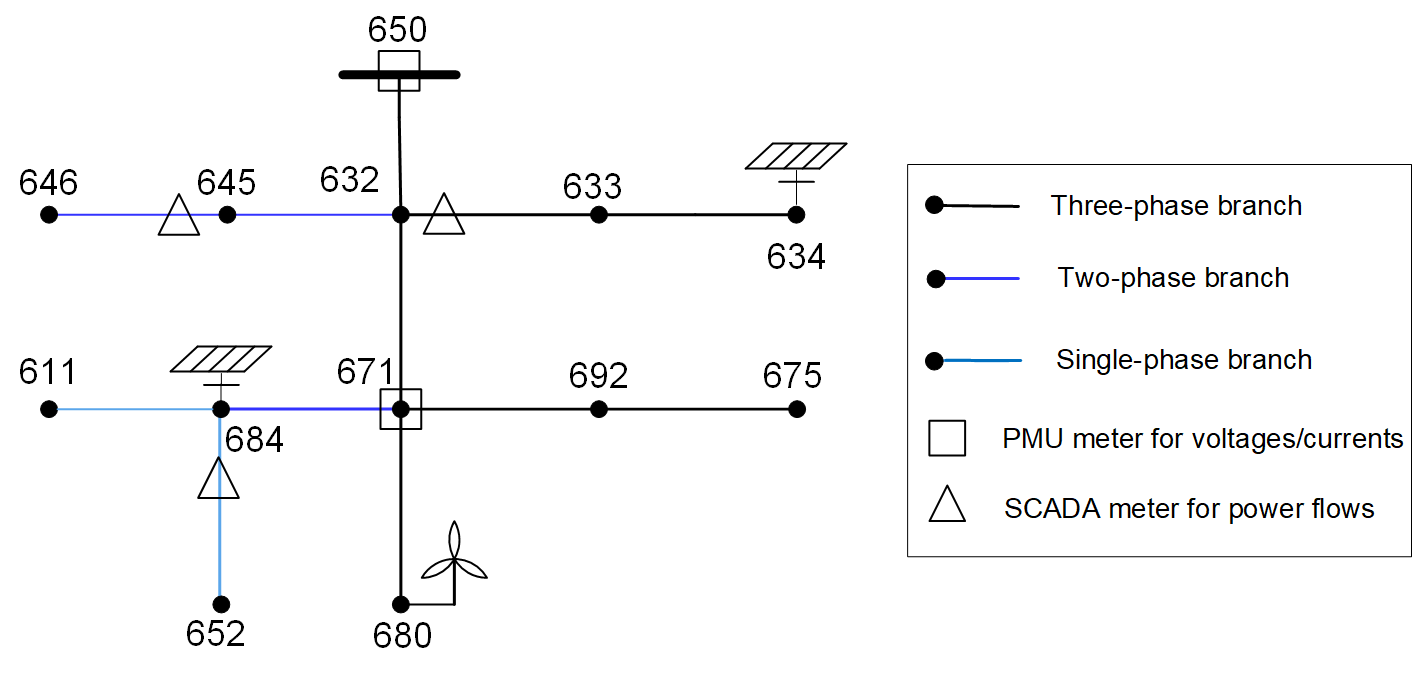}}
\caption{Three-phase unbalanced 13-bus distribution system.}
\label{fig2}
\end{figure}

We test the proposed algorithm on 13- and 34-bus unbalanced IEEE distribution feeders \cite{27ieeetestfeeders}, shown in Fig. 2. The installation details of DER units in the 34-bus test system can be found in \cite{4zhang2020towards}. Real-time PMU and slow-rate SCADA data or pseudo-measurements are collected, and here we suppose that DSSE runs per minute and $T_{s}=10$ min \cite{13song2020dynamic} as an example, and the proposed method can also accommodate other reporting rates. Typical 24-hour trajectories of DER and load profiles are adopted (see them in \cite{13song2020dynamic}). 
Table \ref{Table2_1} lists the placement locations of PMUs and SCADA sensors.
The following maximum errors are adopted in Monte Carlo simulation: 0.1\% for voltage/current magnitudes and $0.1 \mathrm{crad}$ for phase angles in PMU data \cite{5vonmeier2017precision}, 2\% for the SCADA measurements, and 20\% for power injections at all the load/DER nodes as pseudo-measurements \cite{3pau2013efficient}. Then the lagging measurements are generated by adding extra time delay.
\vspace{-2pt}
\subsection{Result Analysis and Computation Time}

\begin{table}[!t]
  \centering
    \caption{Measurement Arrangement in 34-bus Test Systems}
\begin{tabular}{|c|c|c|}
\hline
\multicolumn{2}{|c|}{Measurements} & Location  \\\hline
PMU   & $V$  & 1, 11, 25  \\ \hline
SCADA     &$P_l$, $Q_l$   &1-2, 13-15, 20-23,30-31\\\hline
Pseudo-meas. &$P$, $Q$    &Load and DER nodes       \\\hline
    \end{tabular}%
  \label{Table2_1}%
\end{table}%

\begin{table}[!t]
\centering
    \caption{Comparison of Estimation Accuracy}
\begin{tabular}{|c|c|c|}
\hline 
Average Errors  & MAE [\%] & MAPE [radians] \\
\hline 
\text { \textbf{Proposed Method} } & 0.71 & 0.0062 \\
\text { Extended KF-based DSSE } & 1.52 & 0.0101 \\
\hline 
\end{tabular}
\label{tab2}
\end{table}

We evaluate the estimation accuracy of the proposed method and compare it with an extended KF-based method in \cite{14alcaide2018electric}. 
The latter sets $\alpha=0.6$ and $\beta=0.5$ for all the time snapshots as default. That is, the benchmark method ignores the time skew issue of multi-source multi-rate data in time-varying operating conditions of distribution systems.

Mean absolute percentage errors (MAPEs) for voltage magnitudes and mean absolute errors (MAEs) for phase angles are used to evaluate the estimation accuracy. 
For example, the MAPE of voltages is calculated at node $k$ for each time step: 
\begin{equation}\label{Vm}
\delta_{M, k}^{\varphi}=\frac{1}{n} \sum_{t=1}^n\left|\hat{v}_{k}^{\varphi}-v_{k}^{\varphi}\right| / \left|{v}_{k}^{\varphi}\right| \times 100 \% 
\end{equation}
where $n$ denotes the total number of Monte Carlo simulations.

Table \ref{tab2} lists the MAPEs and MAEs of the proposed method on all nodes in the 34-bus test system, which are the average value on all the nodes. The voltage errors in our method are much lower than the compared method. 
By performing the proposed algorithm, the impacts of these lagging measurements on the estimation voltages are decreased largely.
Because of the existence of lagging measurements and the renewable integration in the feeder, the extended KF-based DSSE method cannot track voltage changes accurately. 

We investigate the computational time in the 13- and 34-bus distribution systems. 
The average CPU time of these two methods is shown in Table \ref{tab3}. The proposed method spends 0.421 seconds in the 34-bus system. It shows that the high computational efficiency of the proposed method makes it promising to track the system states online.

\begin{table}[!t]
\centering
    \caption{Comparison of CPU Time}
\begin{tabular}{|c|c|c|}
\hline 
Average CPU Time at Each Run  & 13-bus System & 34-bus System  \\
\hline 
\text { \textbf{Proposed Method}*} & 0.026 s & 0.421 s \\
\text {Extended KF-based DSSE } & 0.024 s & 0.442 s \\
\hline 
\multicolumn{3}{l}{*The time includes that for online parameter identification.}\\
\end{tabular}
\label{tab3}
\end{table}

\subsection{Adaptive Performance of Proposed DRL-Enabled Method}
We evaluate the adaptive performance of the proposed algorithm in time-varying operating conditions.
In offline training, various operation conditions with $90 \%$ to $110 \%$ of random fluctuations \cite{26zhang2021deep} of these loads/DERs are randomly generated. 
The learning process for the DRL-enabled procedure in the 13-bus system is depicted in Fig. 3. 
Then, the well-trained NNs are adopted online in the identification step to provide the smoothing parameters adaptively. Due to the page limit, here we show the partial estimated voltages by using the proposed adaptive method and \cite{14alcaide2018electric}. Fig.4 compares the estimated trajectories of voltages by these two algorithms at 0:00-1:00. 

Moreover, the three-phase MAEs of the proposed method on all the time steps in Fig.4 are calculated by \eqref{Vm} and compared. It is shown the estimation errors of the B-phase voltages decrease largely to 0.51\%. In contrast, the existing method for this comparison cannot consistently derive the desired accuracy of the voltage estimation in varying operating conditions. The comparison illustrates that the proposed adaptive method further decreases the deviation of the estimated voltages from the true values. To sum up, by utilizing the historical multi-rate measurement data, the proposed adaptive FASE method enhances the state estimation performance.

\begin{figure}[!t]
\centerline{\includegraphics[width=3.2in]{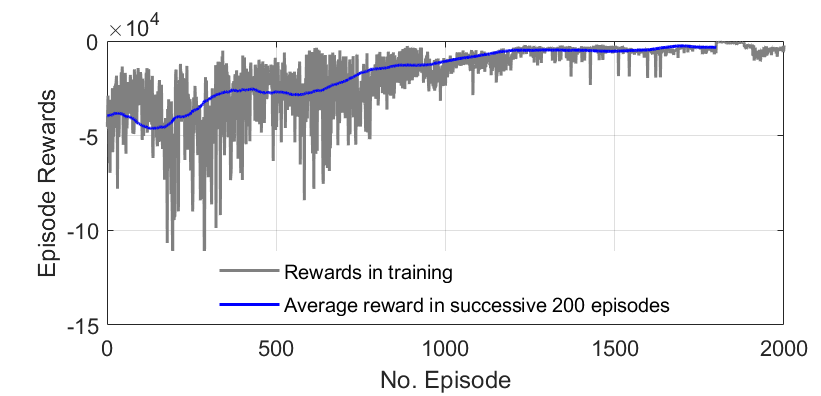}}
\caption{DRL learning process for parameter identification.}
\label{fig}
\end{figure}

\section{Conclusion}
\begin{figure}[!t]
\centerline{\includegraphics[width=3.0in]{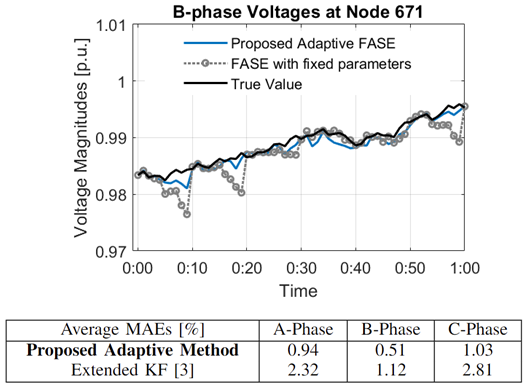}}
\caption{The comparison of estimation errors in multiple timesteps.}
\label{fig4}
\end{figure}


This paper proposes a DRL-enabled adaptive FASE method with multi-rate measurements from various sources in distribution systems. 
The data-driven DRL-enabled framework, which is embedded in the formulation of FASE, realizes adaptive parameter identification in the prediction step. On its online implementation, the proposed method further enables accurate state estimation and tackles the impact of the multi-source multi-rate measurements. In contrast to the existing algorithms with multi-rate measurements, our method tracks the state variation with the varying operational environment efficiently.
\vspace{-6pt}
\bibliographystyle{IEEEtran}
\bibliography{IEEEabrv,Citation}
\let\mybibitem\bibitem

\end{document}